\documentclass{pos}
\usepackage{amssymb,epsfig,amscd}
\title{Brief report on `Radiative $\phi$ decays with derivative interactions'}

\ShortTitle{Radiative $\phi$ decays with derivative interactions}

\author{Francesco Giacosa\\
\thanks{poster session}
        Institute f\"ur Theoretische Physik, Goethe Universit\"at, Max-von-Laue-Str. 1, 60438 Frankfurt am Main, Germany \\
        E-mail: \email{giacosa@th.physik.uni-frankfurt.de}}

\author{Giuseppe Pagliara\\
        \\ Institut  f\"ur Theoretische Physik, Goethe Universit\"at, Max-von-Laue-Str. 1, 60438 Frankfurt am Main, Germany \\
        E-mail: \email{pagliara@th.physik.uni-frankfurt.de}}

\abstract{We study the line shapes of radiative $\phi$-decays with
a direct coupling of the $\phi$ meson to the
$f_{0}(980)$ and $a_{0}(980)$ scalar mesons. The latter couple via derivative interactions 
to $\pi_0 \pi_0$ and $\pi_0 \eta$,  respectively. 
Although the kaon-loop mechanism is usually regarded as the dominant mechanism 
in radiative $\phi$ decays, here we test a different possibility: we set the kaon-loop 
to zero and we fit the theoretical curves to the data by retaining only the direct coupling. 
Remarkably, satisfactory fits can be achieved, mainly due to the effects of 
derivative interactions of scalar with pseudoscalar mesons.}

\FullConference{8th Conference Quark Confinement and the Hadron Spectrum \\
		 September 1-6 2008\\
		 Mainz, Germany}

\begin{document}
The general Lagrangian describing the interactions of scalar and
pseudoscalar mesons reads\cite{ecker}:%

\begin{equation}
\mathcal{L}_{int,f_{0}}=c_{f_{0}\pi\pi}f_{0}\left(  \partial_{\mu
}\overrightarrow{\pi}\right)  ^{2}+d_{f_{0}\pi\pi}M_{\pi}^{2}f_{0}%
\overrightarrow{\pi}^{2}+c_{f_{0}KK}f_{0}\left(  \partial_{\mu}K^{+}\right)
\left(  \partial^{\mu}K^{-}\right)  +d_{f_{0}KK}M_{K}^{2}f_{0}(K^{+}K^{-})+...
\label{f01}%
\end{equation}
This is valid independently on the substructure of the enigmatic light scalar
states. In the chiral limit ($M_{\pi}\rightarrow0,$ $M_{K}\rightarrow0$) only
the derivative interactions survive, but in general also non-derivative
interactions appear.

The radiative decay $\phi\rightarrow f_{0}(980)\gamma\rightarrow\pi^{0}\pi
^{0}\gamma$ (and similarly $\phi\rightarrow a_{0}(980)\gamma\rightarrow\pi
^{0}\eta\gamma$) can occur by the two mechanisms depicted in Fig. 1: (a) the vector
meson couples, via a point-like interaction with coupling $c_{\phi f_{0}%
\gamma}$, to the scalar meson and to the photon. The scalar meson then decays
into pseudoscalars via derivative and non derivative interactions as
preliminarily studied in \cite{black}. (b) The vector meson couples strongly
to kaons. Then, via a kaon-loop a photon and the scalar mesons are generated.%

\begin{figure}
[ptb]
\begin{center}
\includegraphics[
height=1.9571in,
width=4.0231in
]%
{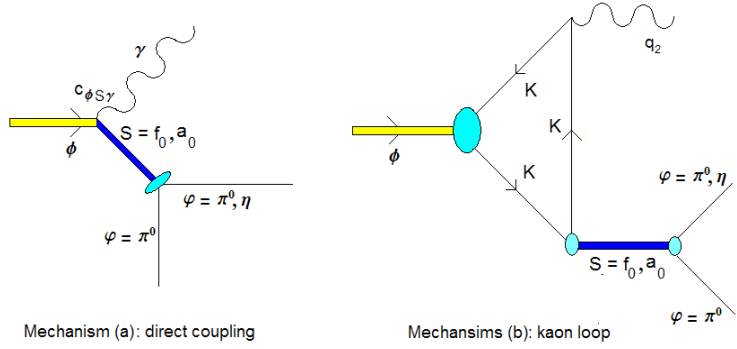}%
\caption{Different mechanisms contributing to the radiative $\phi$ decays.}%
\end{center}
\end{figure}

The decay mode (b), widely studied in the literature, is usually 
considered as the dominant mechanism which contributes to the radiative decays
$\phi\rightarrow S\gamma$ with $S=f_{0},a_{0}$ \cite{achasov} (see also
\cite{oset} and Refs. therein). In Ref. \cite{our}, following the formalism
developed in Ref. \cite{nostro} extended to derivative interactions, we
investigated a different point of view: we considered the possibility that, in agreement with large $N_{c}$ expansion, 
the mechanism (a) is dominant.
Moreover, we set in Eq. (1) $d_{f_{0}\pi\pi}=d_{f_{0}KK}=0$ (and so for
$a_{0}$), thus restricting the study to the chiral limit scenario, in which the
scalars couple \emph{only} via derivatives to the pseudoscalar mesons. The
question we aimed to answer is: \textquotedblleft can one in this scenario fit
the line shapes $\phi\rightarrow\pi^{0}\pi^{0}\gamma$ and $\phi\rightarrow
\pi^{0}\eta\gamma$ as measured in the SND and KLOE experiments
\cite{achasov2,aloisio}?\textquotedblright\ Quite remarkably, as shown in Fig.
(2), the answer is positive (see Ref. \cite{our} for details). Some remarks are in order:

\begin{figure}[ptb]
\hbox{\hskip -0.5cm \epsfig{file=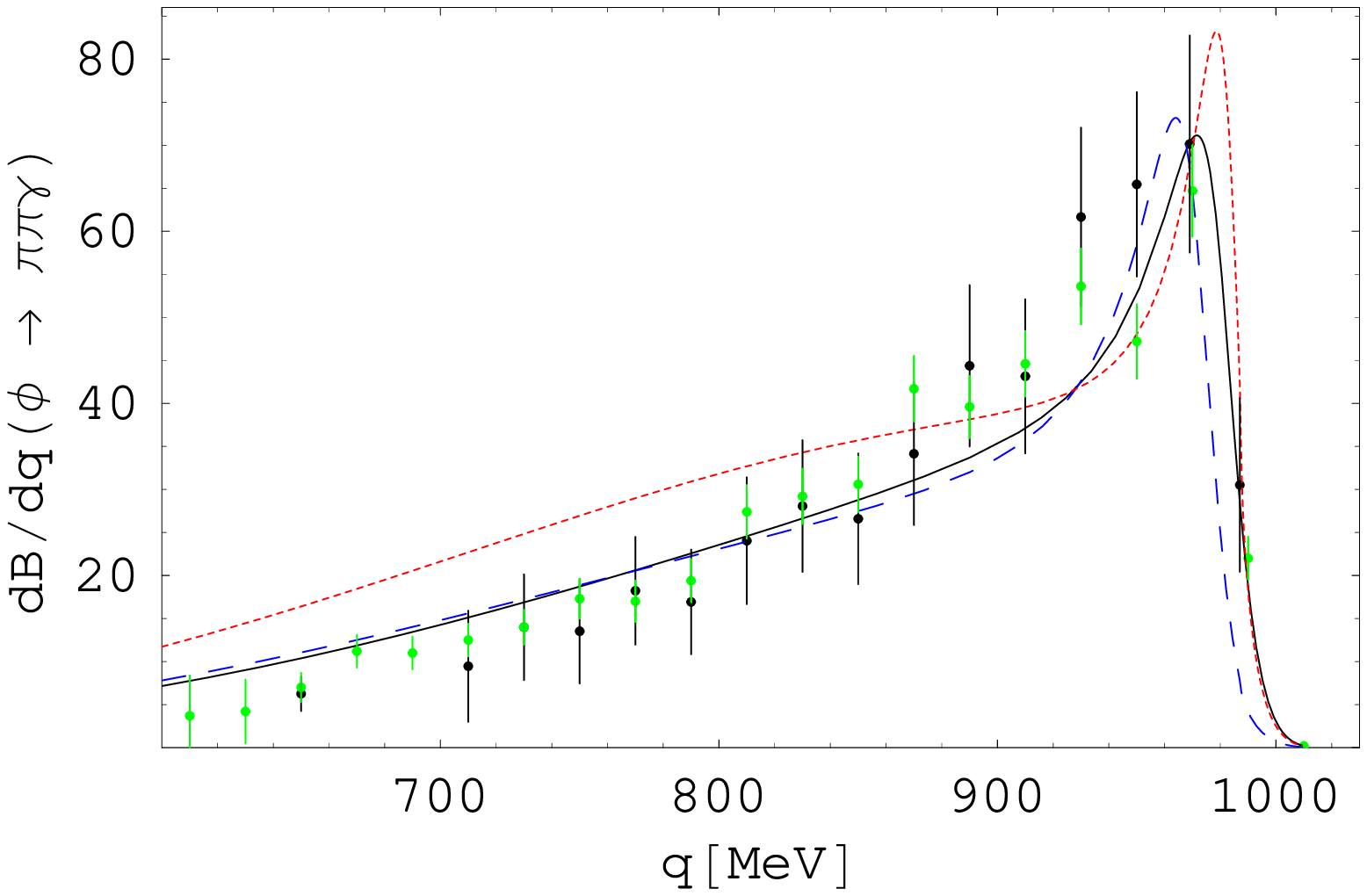,height=4.5cm}\hskip 1.cm}
\vskip -4.7cm
\hbox{\hskip 7cm \epsfig{file=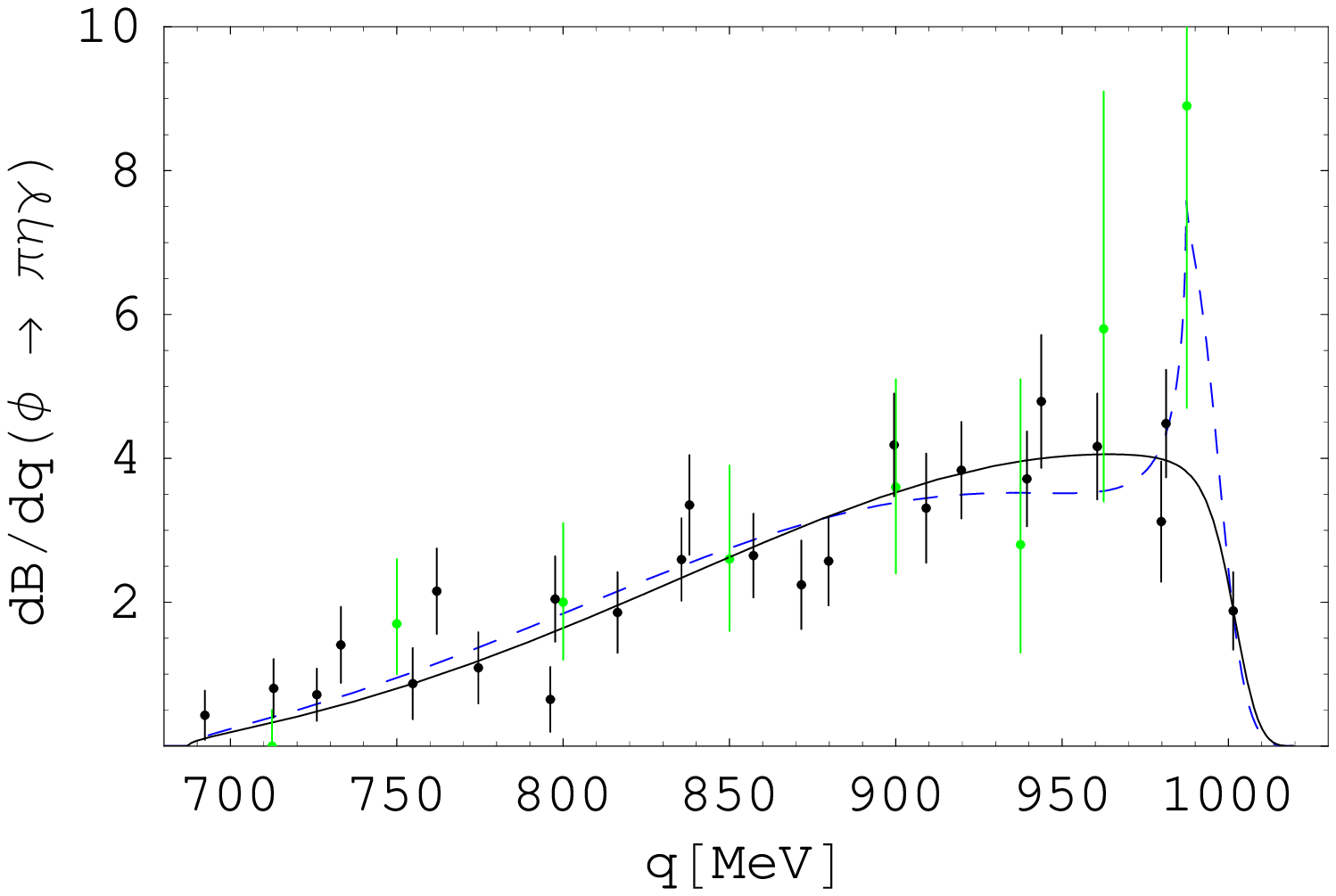,height=4.6cm}\hskip 1.cm }\vskip 0cm\caption{Left
panel: Branching ratio $dB(\phi\rightarrow\pi^{0}\pi^{0}\gamma)/dq\cdot10^{8}$
$MeV^{-1}$ where $q$ is the invariant $\pi^{0}\pi^{0}$ mass. We consider data
sets from the SND and KLOE collaborations \cite{achasov2,aloisio}
corresponding respectively to the black and green dots. The continuous line is
the result of the fit by setting $c_{f_{0}KK}=0,$ the dashed blue line
corresponds to the case $c_{f_{0}KK}=12$ GeV$^{-1}$. The dotted red line
corresponds also to $c_{f_{0}KK}=12$ GeV$^{-1}$ but only data points above 0.8
GeV are used in the fit. Right panel: $dB(\phi\rightarrow\pi^{0}\eta
\gamma)/dq\times10^{7}(MeV^{-1}).$ Green points from \cite{achasov2} and black
ones from \cite{aloisio}. The solid line corresponds to $c_{a_{0}KK}=0$ while
the dashed blue one -with the pronounced peak at threshold- to $c_{a_{0}%
KK}=12$ GeV$^{-1}$.}%
\end{figure}

(i) When mechanism (b) is dominant, the amplitude of the process is
proportional to the coupling of the scalar mesons to the pseudoscalars (as,
for instance, $c_{f_{0}KK}$ in Eq. (1)). However, when mechanism (a) is
dominant,  the coupling
of the scalars to kaons appears only in the denominator of the scalar
particles, see Fig. 1. Thus, it is more difficult to extract them from experiment .

(ii) Fits in the $f_{0}$ channel in the region between 0.6 and 1 GeV are
acceptable, see Fig. 2, left panel. However, when the coupling to kaons is
large, as suggested in Ref. \cite{buggo1}, the quality of the fit decreases.
Alternatively, one can satisfactory fits the data between 0.8 and 1 GeV, but then
the theoretical curve overshoots  data points below 0.8 GeV. This suggests the existence of the
$\sigma\equiv f_{0}(600)$ meson, which is expected to generate a destructive
interference with the $f_{0}$ meson in a similar way as presented in Ref.
\cite{aloisio}.

(iii) Fits in the $a_{0}$ channel for different values of the coupling
$c_{a_{0}KK}$ are good  (Fig.2 , right panel) . Interestingly, when the latter is large, 
as found in Ref. \cite{buggo2}, a very narrow peak is obtained
close to threshold.

In summary, the results of the present study point out that derivative
scalar-to-pseudoscalar interactions, being tailor-made to reproduce peaks
close to threshold, can play an important role in the description of radiative
$\phi$ decays. If, on the contrary, non-derivative
interactions are used, worse fits are obtained. Future studies will show which one of  the mechanisms of Fig. 1
(if any) is dominant.

\end{document}